\documentclass[aps,twocolumn,prd,showpacs,showkeys,preprintnumbers,superscriptaddress,nobibnotes,floatfix,longbibliography,nofootinbib]{revtex4-2}
\pdfoutput=1
\usepackage{amsmath}
\usepackage{amsfonts}
\usepackage{amssymb}
\usepackage{mathrsfs}
\usepackage{graphicx}
\usepackage{color}
\usepackage{hyperref}
\usepackage{multirow}
\usepackage{slashed}
\usepackage{epsf}

\usepackage{graphicx}
\usepackage{caption}
\usepackage{subcaption}

\usepackage{color}

\def\tl{\tilde l}

\def\tl{\tilde l}

\def\tl{\tilde}

\begin{document}

\title{
The Right-Handed Slepton Bulk Regions for Dark Matter in the Generalized Minimal Supergravity (GmSUGRA)
}

\author{Imtiaz Khan}
\affiliation{CAS Key Laboratory of Theoretical Physics, Institute of Theoretical Physics, Chinese Academy of Sciences, Beijing 100190, China}
\affiliation{School of Physical Sciences, University of Chinese Academy of Sciences, No. 19A Yuquan Road, Beijing 100049, China}

\author{Waqas Ahmed}
\affiliation{Center for Fundamental Physics and School of Mathematics and Physics, Hubei Polytechnic University}

\author{Tianjun Li}
%\email{tli@mail.itp.ac.cn}
\affiliation{CAS Key Laboratory of Theoretical Physics, Institute of Theoretical Physics, Chinese Academy of Sciences, Beijing 100190, China}
\affiliation{School of Physical Sciences, University of Chinese Academy of Sciences, No. 19A Yuquan Road, Beijing 100049, China}

\author{Shabbar Raza}
%\email{shabbar.raza@fuuast.edu.pk}
\affiliation{Department of Physics, Federal Urdu University of Arts, Science and Technology, Karachi 75300, Pakistan}

%\date{\today}
%%%%%%%%%%%%%%%%%%%%%%%%%%%%%%%%%%%%%%%%%%%%%%%%%%%%%%%%%%%%%%%%%%%%%%%
\begin{abstract}

We study the light right-handed slepton bulk regions for dark matter from the Generalized Minimal Supergravity (GmSUGRA) in the Minimal Supersymmetric Standard Model (MSSM). In our comprehensive numerical studies, we show that $\mathcal{R_{\tilde{\phi}}}\gtrsim10\%$ is a conservative criteria to formulate bulk region, where $\mathcal{R_{\tilde{\phi}}}\equiv({m_{\tilde{\phi}}-m_{\tilde{\chi}_1^0}})/{m_{\tilde{\chi}_1^0}}$. 
For right-handed stau as the Next to the Lightest Supersymmetric Partcile (NLSP), we find a large viable parameter space, consistent with the current LHC constraints, Planck2018 dark matter relic density bounds, and direct bounds on neutralino-nucleons scattering cross-section that naturally supports the right-handed stau bulk regions for dark matter. In particular, the upper bounds on the masses of the Lightest Supersymmetric Particle (LSP) neutralino and right-handed stau are about 120.4 GeV and 138 GeV, respectively.
 This bulk region may be beyond the current LHC reach and could be probed at LUX-ZEPLIN, a next-generation dark matter direct detection experiment, the Future Circular Collider (FCC-ee) at CERN, and the Circular Electron Positron Collider (CEPC).
 However, the scenario with the right-handed selectron as the NLSP is excluded by the LHC supersymmetry searches.

\end{abstract}
%\pacs{}
\maketitle

%%%%%%%%%%%%%%%%%%%%%%%%%%%%%%%%%%%%%%%%%%%%%%%%%%%%%%%%%%%%%%%%%%%%%%%

\section{Introduction}
\label{intro}
 Even though supersymmetric standard models (SSMs) are the most promising candidate to explain physics beyond the standard model (BSM), experimentally, no concrete evidence has been found so far. In the SSMs, we can achieve gauge couplings unification\cite{gaugeunification,Georgi:1974sy,Pati:1974yy,Mohapatra:1974hk,Fritzsch:1974nn,Georgi:1974my}, with R-Parity, the lightest supersymmetric particle (LSP) as a good cold dark matter (DM) candidate \cite{neutralinodarkmatter,darkmatterreviews}, a natural solution to the gauge hierarchy problem, and break the electroweak (EW) gauge symmetry radiatively because of the large top quark Yukawa coupling. Besides, the minimal SSM (MSSM) predicts the Higgs mass [100,135] GeV~\cite{mhiggs}. Thus, supersymmetry (SUSY) has been one of the main focuses of the searches being done in the Large Hadron Collider (LHC) to look for the BSM, provides a bridge between the low-energy phenomenology and high-energy fundamental physics, and so the most promising candidate for the new physics beyond the SM.
 
 However, after the LHC Run-2, we still did not have any sign of SUSY, and LHC searches established strong constraints on the SSMs spectra. The searches have been done so far elevated low mass bounds on the masses of gluino, first-two generation squarks, stop, and sbottom to around 2.2~TeV, 1.9~TeV, 1.25~TeV, and 1.5~TeV, respectively~\cite{ATLAS-SUSY-Search, Aad:2020sgw, Aad:2019pfy, CMS-SUSY-Search-I, CMS-SUSY-Search-II}. Thus, at least the colored supersymmetric particles (sparticles) must be heavy around a few TeV. If the LSP is a bino-like neutralino, then the large sfermions masses lead to a small annihilation cross-section, as the process mediated by sfermions is suppressed. Thus, in the absence of some other annihilation-enhancement or the coannihilation mechanism, the resulting DM relic abundance would be far in excess of the value measured by the Planck $(5\sigma)$ bound, $0.114 \leq \Omega_{\rm CDM}h^2 (\rm Planck) \leq 0.126$~\cite{Akrami:2018vks}\cite{Drees:1992am}. Despite these constraints on the squark masses, the current data still leave open the possibility of models for an electroweak-scale bino dominant LSP and relatively light sleptons with much heavier squarks. SUSY models inspired by the BSM indicate light sleptons mass around a few hundred GeV up to TeV scale, for example, see \cite{Ahmed:2022ude} and the references therein. However, in the standard scenario for the bino dominant LSP, there are difficulties with obtaining a large enough annihilation cross-section to deplete the DM relic density, as the bino DM always annihilates through the process $\tilde{\chi}_1^0 \tilde{\chi}_1^0\rightarrow f \bar{f}$ via t and u-channel sfermions exchange where $f, \bar{f}$ are standard model fermions and this process is not sufficient because of the intermediate sfermions large masses.  However, the four distinct approaches to the observed DM relic density: (a) Bulk region where sfermions are light; (b) The Z/Higgs resonance or Z/Higgs funnel where the neutralino LSP mass is about half of the masses of the CP-even Higgs $H_0$, CP-odd Higgs $A_0$, SM Higgs, or Z bosons; (c) Coannihilation, where the sfermion masses are closed to the LSP neutralino; or (d) Mixing scenario or well-tempered scenario, where the LSP neutralino has wino or higgsino component to significant increase the annihilation cross-section, the light sfermions exchange or the bulk region is the most natural version of neutralino DM, wherein no coannihilaiton or resonance annihilation mechanism is necessary to suppress the relic abundance\cite{King:2006tf}. Thus the region of parameter space where this works out right is often referred to by the jargon ``bulk region''. To escape the LHC SUSY search constraints and to be consistent with various experimental results, some of us proposed the Electroweak Supersymmetry (EWSUSY)~\cite{Cheng:2012np, Cheng:2013hna, Li:2014dna}, where the squarks and gluinos are around a few TeV while the sleptons, sneutrinos, bino, and winos are within one TeV. The higgsinos (or say the Higgs bilinear $\mu$ term) can be either heavy or light. Especially, the EWSUSY can be realized in the Generalized Minimal Supergravity (GmSUGRA)~\cite{Li:2010xr, Balazs:2010ha}. This article discusses a bulk region in the MSSM from the GmSUGRA. In order to uncover the bulk region in the MSSM via GmSUGRA, we can only consider the right-light sleptons with all other sfermions must be heavily indicated by the LHC SUSY searches. To determine whether the interaction between sfermions and the LSP is annihilation or coannihilation, the mass difference between the light right-handed sfermions and LSP is important i.e, the ratio of the mass difference $\mathcal{R_{\tilde{\phi}}}\equiv({m_{\tilde{\phi}}-m_{\tilde{\chi}_1^0}})/{m_{\tilde{\chi}_1^0}}$ is important, where $\tilde{\phi}$ is $\tilde {e}_R$ (right-handed light selection) or $\tilde{\tau}_1$ (light stau). The $\mathcal{R_{\tilde{\phi}}}\geq 10\%$ is the conservative criteria to observe DM relic density solely via annihilation, and not from coannihilation or resonance, etc. 
 For right-handed stau as the Next to the Lightest Supersymmetric Partcile (NLSP), we find a large viable parameter space, consistent with the current LHC constraints, Planck2018 dark matter relic density bounds, and direct bounds on neutralino-nucleons scattering cross-section that naturally supports the right-handed stau bulk regions for dark matter. In particular, the upper bounds on the masses of the Lightest Supersymmetric Particle (LSP) neutralino and right-handed stau are about 120.4 GeV and 138 GeV, respectively.
 This bulk region may be beyond the current LHC reach and could be probed at LUX-ZEPLIN, a next-generation dark matter direct detection experiment, the Future Circular Collider (FCC-ee)~\cite{FCC:2018byv,FCC:2018evy} at CERN, and the Circular Electron Positron Collider (CEPC)~\cite{CEPCStudyGroup:2018ghi}.
 However, the scenario with the right-handed selectron as the NLSP is excluded by the LHC supersymmetry searches.

This paper is organized as follows. In section \ref{model}, we briefly summarize the model and relevant free parameters. In section \ref{scanning}, we describe the scanning procedure and range of our GUT scale parameters, and in section \ref{constraints}, we highlight the phenomenological constraints. In section \ref{bulk}, we explain the numerical results and finally, we conclude our findings in section \ref{conclusion}.

%%%%%%%%%%%%%%%%%%%%%%%%%%%%%%%%%%%%%%%%%%Model Part%%%%%%%%%%%%%%%%%%%%%
\section{The EWSUSY from the GmSUGRA in the MSSM}
\label{model}
The EWSUSY can be realized in the GmSUGRA~\cite{Li:2010xr, Balazs:2010ha}. As stated in~\cite{Cheng:2012np,Cheng:2013hna, Li:2014dna}, in this framework, the sleptons and electroweakinos (charginos, bino, wino, and/or higgsinos) are within one TeV while squarks and/or gluinos 
can be in several TeV mass ranges~\cite{Li:2014dna,Cheng:2012np}. Apart from this, the gauge coupling relation and gaugino mass relation at the GUT scale are ~\cite{Li:2010xr, Balazs:2010ha},
\begin{equation}
 \frac{1}{\alpha_2}-\frac{1}{\alpha_3} =
 k~\left(\frac{1}{\alpha_1} - \frac{1}{\alpha_3}\right)~,
\end{equation}
\begin{equation}
 \frac{M_2}{\alpha_2}-\frac{M_3}{\alpha_3} =
 k~\left(\frac{M_1}{\alpha_1} - \frac{M_3}{\alpha_3}\right)~,
\end{equation}
where $k$ is the index of these relations and is equal to 5/3 in our simple GmSUGRA. We assume for simplicity that at the GUT scale ($\alpha_1=\alpha_2=\alpha_3$), the gaugino mass relation becomes
\begin{equation}
 M_2-M_3 = \frac{5}{3}~(M_1-M_3)~,
\label{M3a}
\end{equation}
 It is obvious that
the universal gaugino mass relation $M_1 = M_2 = M_3$ in the mSUGRA, is just a special case of 
a general one, that's why we called it GmSUGRA. In this case, there are two independent gauginos rather than three. Thus, Eq.~(\ref{M3a}) implies for $M_2$ in terms of $M_1$ and $M_3$ as free input parameters is as follow:
\begin{eqnarray}
M_2=\frac{5}{3}~M_1-\frac{2}{3}~M_3~.
\label{M3}
\end{eqnarray}

We use Ref.~\cite{Balazs:2010ha} for the general SSB GUT scale scalar masses. The masses of squarks obtained in the $SU(5)$ model with an adjoint Higgs field, where we employ Slepton masses as a free parameter.
\begin{eqnarray}
m_{\tl{Q}_i}^2 &=& \frac{5}{6} (m_0^{U})^2 +  \frac{1}{6} m_{\tl{E}_i^c}^2~,\\
m_{\tl{U}_i^c}^2 &=& \frac{5}{3}(m_0^{U})^2 -\frac{2}{3} m_{\tl{E}_i^c}^2~,\\
m_{\tl{D}_i^c}^2 &=& \frac{5}{3}(m_0^{U})^2 -\frac{2}{3} m_{\tl{L}_i}^2~.
\label{squarks_masses}
\end{eqnarray}
Here, $m_{\tl Q}$, $m_{\tl U^c}$, $m_{\tl D^c}$, $m_{\tl L}$, and  $m_{\tl E^c}$ represent the left-handed scalar squark doublets, right-handed up-type squarks, right-handed down-type squarks, left-handed sleptons, and right-handed sleptons, respectively, and $m_0^U$ is the universal scalar mass, as in the mSUGRA. For the light sleptons in the EWSUSY, $m_{\tl L}$ and $m_{\tl E^c}$ are both being within 1 TeV. Especially, in the limit $m_0^U \gg m_{\tl L/\tl E^c}$, we get the approximated relations for squark masses: $2 m_{\tl Q}^2 \sim m_{\tl U^c}^2 \sim m_{\tl D^c}^2$. In addition, the Higgs soft masses $m_{\tl H_u}$ and $m_{\tl H_d}$, and the  trilinear soft terms $A_U$, $A_D$ and $A_E$ can all be free parameters from the GmSUGRA~\cite{Cheng:2012np,Balazs:2010ha}.

\section{Scanning procedure and GUT scale parameters range}
\label{scanning}
We employ the ISAJET~7.85 package~\cite{ISAJET} to carry out a random scan over the parameter space given below. In this package, the MSSM renormalization group equations (RGEs) in the $\overline{DR}$ regularization scheme are evolved for the third-generation Yukawa couplings from the weak scale to $M_{\rm GUT}$ value. We do not enforce the unification condition $g_3=g_1=g_2$ at $M_{\rm GUT}$ (where $g_{3}$,$g_2$ and $g_1$ are the $SU(3)_{C}$, $SU(2)_{L}$ and $U(1)_{Y}$ gauge couplings) strictly since a few percent variations from unification can be allotted to the unknown GUT-scale threshold corrections~\cite{Hisano:1992jj}. All the SSB parameters, along with the gauge and Yukawa couplings, are evolved back to the weak scale $M_{\rm Z}$, with the boundary conditions given at $M_{\rm GUT}$ (for more detail see \cite{ISAJET}). ISAJET employs two-loop MSSM renormalization group equations (RGEs) and defines $M_{\rm U}$ to be the scale at which $g_1=g_2$. Using the parameters discussed in section \ref{model}, we have performed the random scans for the following parameter ranges
\begin{align}
100 \, \rm{GeV} \leq & m_0^{U}  \leq 10000 \, \rm{GeV}  ~,~\nonumber \\
0 \, \rm{GeV} \leq & M_1  \leq 1200 \, \rm{GeV} ~,~\nonumber \\
1000\, \rm{GeV} \leq & M_3   \leq 3000 \, \rm{GeV} ~,~\nonumber \\
100 \, \rm{GeV} \leq & m_{\tilde L}  \leq 5000 \, \rm{GeV} ~,~\nonumber \\
0 \, \rm{GeV} \leq & m_{\tilde E^c}  \leq 300 \, \rm{GeV} ~,~\nonumber \\
0 \, \rm{GeV} \leq & m_{\tilde H_{u,d}} \leq 10000 \, \rm{GeV} ~,~\nonumber \\
-10000 \, \rm{GeV} \leq & A_{U}=A_{D} \leq 10000 \, \rm{GeV} ~,~\nonumber \\
-5000 \, \rm{GeV} \leq & A_{E} \leq 5000 \, \rm{GeV} ~,~\nonumber \\
2\leq & \tan\beta  \leq 60~.~
 \label{input_param_range}
\end{align}
Also, we consider  $\mu > 0$ and  use $m_t = 173.3\, {\rm GeV}$  \cite{:2009ec}.
Note that our results are not too sensitive to one
 or two sigma variations in the value of $m_t$  \cite{bartol2}. Note, we will use the notations $A_{U}, A_D$ and $A_E$, for $A_t, A_b, A_{\tau}$ receptively.
 In scanning the parameter space, we employ the Metropolis-Hastings
 algorithm as described in \cite{Belanger:2009ti}.
The data points collected are Radiative Electroweak Symmetry Breaking (REWSB) compatible
 with the neutralino being the LSP.

\section{phenomenological Constraints}
\label{constraints}
 The data points collected are the REWSB-compatible, with the neutralino serving as the LSP. Besides, we impose the bounds that the LEP2 experiments set on charged sparticle masses ($\gtrsim 100$ GeV) \cite{Patrignani:2016xqp}, for Higgs mass bounds \cite{Khachatryan:2016vau} 
 due to the theoretical uncertainty in the calculation of $m_h$ in the MSSM -- see {\it e.g.}~\cite{Allanach:2004rh} --
we apply the constraint for the Higgs boson mass to our results as $m_{h}=[122,128] {\rm GeV}$. In addition, based on~\cite{ATLAS-SUSY-Search, Aad:2020sgw, Aad:2019pfy, CMS-SUSY-Search-I, CMS-SUSY-Search-II}, we consider the constraints on gluino and first/second generation of squarks masses $m_{\widetilde g} \gtrsim ~ 2.2 \,{\rm TeV}$, and $m_{\widetilde q} \gtrsim ~ 2 \,{\rm TeV}$ respectively.  We also consider the constraints from rare decay processes $B_{s}\rightarrow \mu^{+}\mu^{-} $ \cite{Aaij:2012nna}, $b\rightarrow s \gamma$ \cite{Amhis:2012bh}, and $B_{u}\rightarrow \tau\nu_{\tau}$ \cite{Asner:2010qj}. Besides, we require the relic abundance of the LSP neutralino to satisfy the Planck2018 bound \cite{Akrami:2018vks}. More explicitly, we set
\begin{eqnarray}
m_h  = 122-128~{\rm GeV}~~&
\\
m_{\tilde{g}}\geq 2.2~{\rm TeV}, m_{\widetilde q} \gtrsim ~ 2 \,{\rm TeV} ~~  \\
0.8\times 10^{-9} \leq{\rm BR}(B_s \rightarrow \mu^+ \mu^-)
  \leq 6.2 \times10^{-9} \;(2\sigma)~~&&
\\
2.99 \times 10^{-4} \leq
  {\rm BR}(b \rightarrow s \gamma)
  \leq 3.87 \times 10^{-4} \; (2\sigma)~~&&
\\
0.15 \leq \frac{
 {\rm BR}(B_u\rightarrow\tau \nu_{\tau})_{\rm MSSM}}
 {{\rm BR}(B_u\rightarrow \tau \nu_{\tau})_{\rm SM}}
        \leq 2.41 \; (3\sigma)~~&&
\\
 0.114 \leq \Omega_{\rm CDM}h^2 (\rm Planck) \leq 0.126   \; (5\sigma)~~&&.
%\\
% 2.1 \times 10^{-10} \leq \Delta a_{\mu}
%  \leq 50.1 \times 10^{-10} \; (3\sigma)~~&\cite{BNL}&
%\labels{constraints}
\end{eqnarray}

%%%%%%%%%%%%%%%%%%%%%%%Discussion%%%%%%%%%%%%%%%%%%%%%%%%%%%%%%%%%%%%%%
\section{Numerical results and Discussion}
\label{bulk}
\begin{figure}[h!]
	%\centering \includegraphics[width=7.90cm]{R1total.png}
    \centering \includegraphics[width=7.90cm]{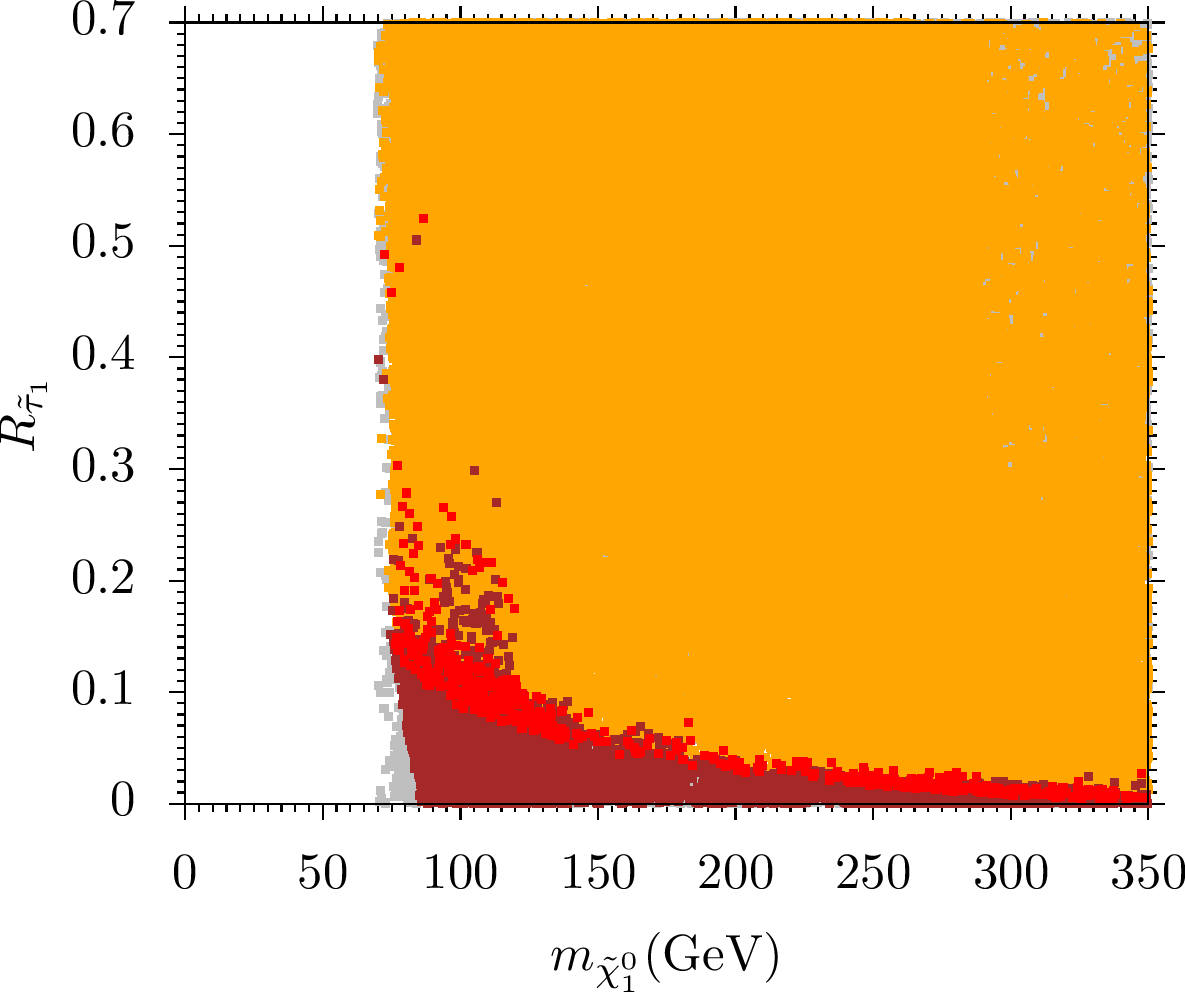}
	\caption{ Grey points satisfy the REWSB and yield LSP neutralino. Orange, brown, and red points are the subset of grey points that satisfy LEP bounds, B-physics bounds, Higgs bound, and sparticles LHC constraints. Also, orange, brown, and red points respectively correspond to over-saturated, under-saturated, and saturated DM relic density. In the panel,$\mathcal{R}_{\tilde{\tau}_1}\equiv({m_{\tilde{\tau}_1}-m_{\tilde{\chi}_1^0}})/{m_{\tilde{\chi}_1^0}}$}.
\label{F1}
\end{figure}

\begin{figure}[h!]
	\centering \includegraphics[width=7.90cm]{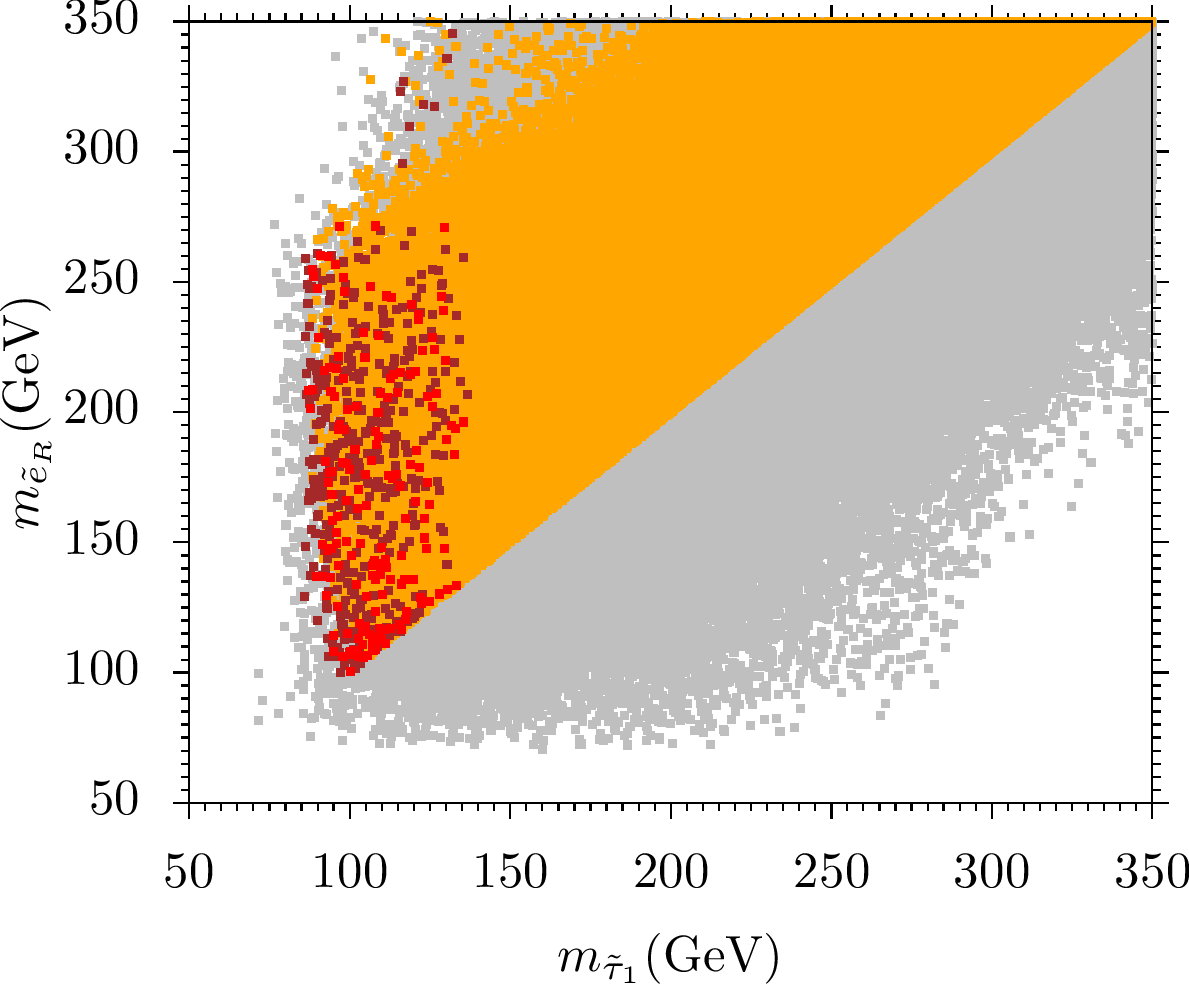}
	\caption{ The penal description and color coding are the same as in figure \ref{F1} with $\mathcal{R}_{\tilde{\tau}_1}\gtrsim10\%$}.
\label{F11}
\end{figure}
\begin{figure}[h!]
	\centering \includegraphics[width=7.90cm]{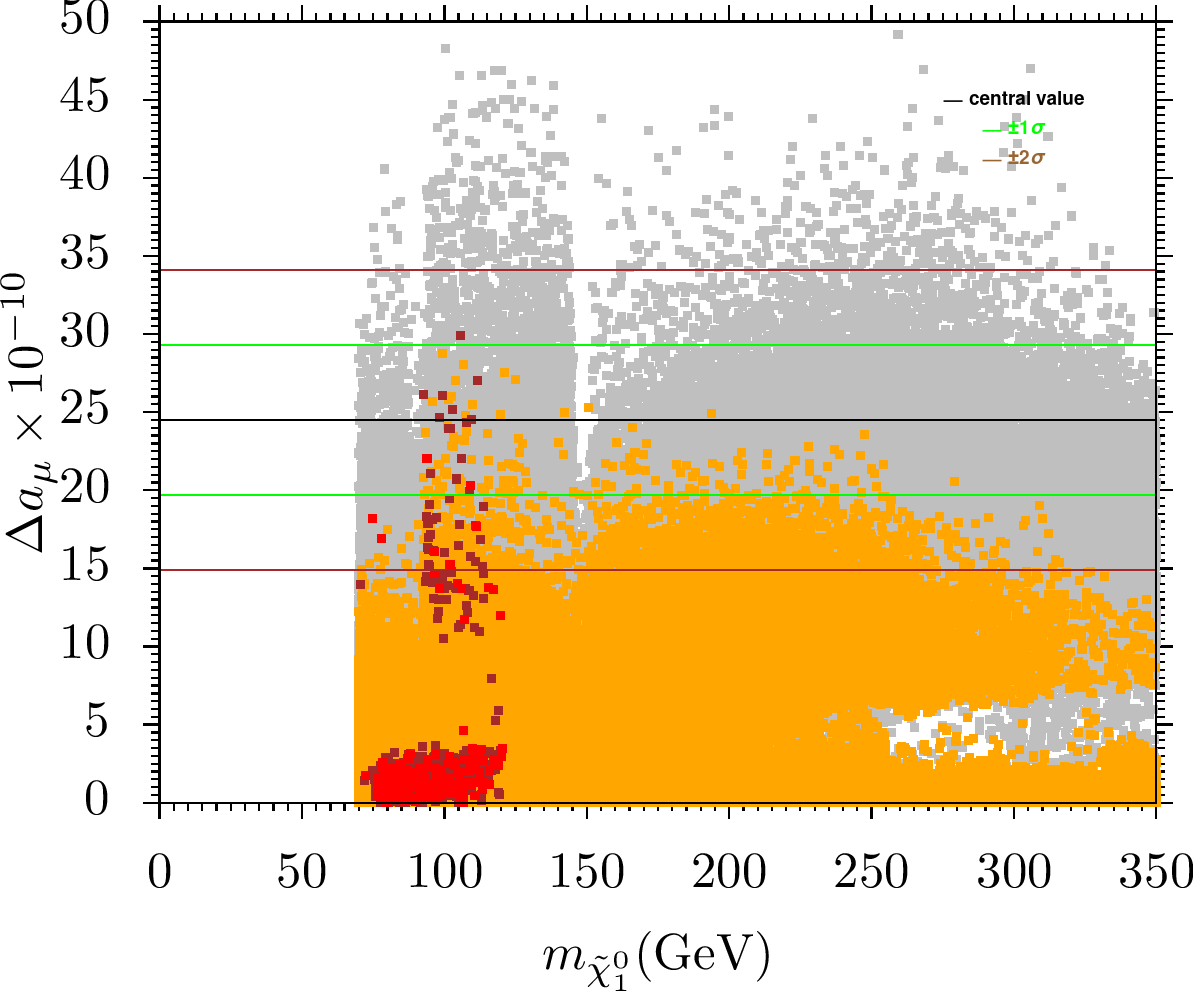}
	\caption{ The penal description and color coding are the same as in figure \ref{F1} with $\mathcal{R}_{\tilde{\tau}_1}\gtrsim10\%$}. The black line shows the central value of $\Delta a_{\mu}$ and the green and brown lines represent $1\sigma$ and $2\sigma$ deviation from the central value.
\label{Fg-2}
\end{figure}

\begin{figure}[h!]
	\centering \includegraphics[width=7.90cm]{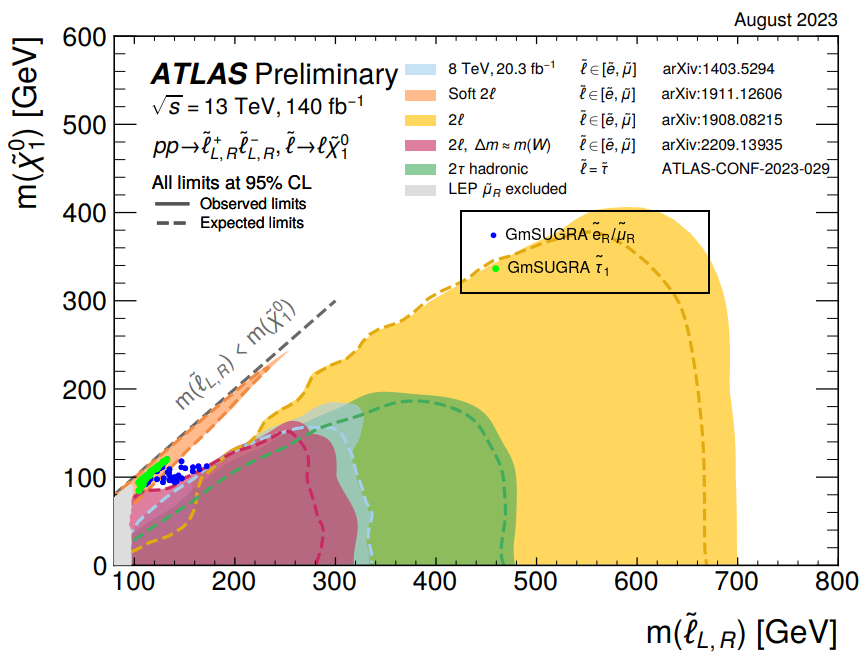}
	\caption{\small {The bulk region from the GmSUGRA superimposed over the August $2023$ ATLAS SUSY updated summary plots \cite{ATLAS:2023xco} for the electroweak production of sleptons \cite{ATLAS:2014zve,ATLAS:2019lng,ATLAS:2019lff,ATLAS:2022hbt,ATLAS:2023djh}. The blue point corresponds to $\tilde{e}_R=\tilde{\mu}_R$ and green points correspond to $\tilde{\tau}_1$ in the GmSUGRA. All points are annihilation only adhere to our requirement $\mathcal{R}_{\tilde{\tau}_1}\gtrsim10\%$. Note that the ATLAS orange shaded region applies to $\tilde{e}_R=\tilde{\mu}_R$ only, not to $\tilde{\tau}_1$. The $\tilde{\tau}_1$ constraints are the green shaded region in the ATLAS, and the GmSUGRA points are comfortably beyond.}}
	\label{F4}
\end{figure}

%%%%%%%%%%%%%%%%%%%%%%%%%%%%%%%%%%%%%%%%%%%%%%%%%%%%%%%%%%%%%%%%%%%%%%%
%%%%%%%%%%%%%%%%%%%%%%%%%%%%%%%%%%%%%%%%%%%%%%%%%%%%%%%%%%%%%%%%%%%%%%%
\begin{table}[h!]
	\centering
	%\begin{ruledtabular}
	\scalebox{0.8}{
		\begin{tabular}{|l|cc|}
			\hline
			\hline
			& Point 1 & Point 2     \\
			\hline
			$m_{0}^{U}$          &   1210      & 1294  \\
			$M_{1},M_{2},M_{3} $         &   281.1,-1449.5,2877      & 267.8, -1287, 2600\\
			$m_{E^c},m_{L}$      &   125.2,1054    &100.6, 1142  \\
			$m_{H_{u}},m_{H_{d}}$           &    3012,901.6     &  3308, 1024 \\
			$m_{Q},m_{U^{c},m_{D^{c}}}$    & 1105.8,1558.8,1303.7 & 1182, 1668.5, 1386.1 \\
			$A_{t}=A_{b},A_{\tau}$            &    -7248,-349.2     & -6868, -332 \\
			$\tan\beta$                      & 21.8 & 19.6\\
			\hline
			$m_h$            &  125    & 125   \\
			$m_H$            &  3545    & 3109 \\
			$m_{A} $         &  3522     &  3088   \\
			$m_{H^{\pm}}$    &  3546   & 3110    \\
			\hline
			$m_{\tilde{\chi}^0_{1,2}}$
			& -111,1291 & -105,1147\\
			$m_{\tilde{\chi}^0_{3,4}}$
			& -3692,3692 &-3164,3165 \\
			$m_{\tilde{\chi}^{\pm}_{1,2}}$
			&1297,3696  & 1152,3169  \\
			\hline
			$m_{\tilde{g}}$  & 5884    &5360 \\
			\hline $m_{ \tilde{u}_{L,R}}$
			& 5222,5269  & 4796,4883   \\
			$m_{\tilde{t}_{1,2}}$
			& 3384, 4319  & 2978,3902 \\
			\hline $m_{ \tilde{d}_{L,R}}$
			& 5223,5205 & 4797,4800\\
			$m_{\tilde{b}_{1,2}}$
			& 4308,4969 & 3892,4602 \\
			\hline
			$m_{\tilde{\nu}_{1}}$
			& 1376       & 1383  \\
			$m_{\tilde{\nu}_{3}}$
			& 1360      & 1370  \\
			\hline
			$m_{ \tilde{e}_{L,R}}$
			& 1395,148   & 1399, 136 \\
			$m_{\tilde{\tau}_{1,2}}$
			& 124,1374    & 117, 1381\\
			\hline
			$\Delta a_{\mu}$
			& 2.4$\times 10^{-10}$ & 1.5$\times 10^{-10}$ 
			\\
			\hline
			$\sigma_{SI}(pb)$
			& 4.6$\times 10^{-14}$ & 1.1$\times 10^{-13}$\\
   $\sigma_{SD}(pb)$
			& 2.6$\times 10^{-10}$ & 5.8$\times 10^{-10}$  
			\\
			$\Omega_{CDM}h^2$
			& 0.125      & 0.118 \\
    $\mathcal{R}_{\tilde{\tau}_1}$& $12\%$   &  $11\%$     \\
    $\mathcal{R}_{\tilde{e}_R}$& $33\%$   &  $28\%$     \\
			\hline
			\hline
		\end{tabular}
  }
	%\end{ruledtabular}
	\caption{All the masses are in the unit of GeV where $\mathcal{R}_{\tilde{\tau}_1}\equiv({m_{\tilde{\tau}_1}-m_{\tilde{\chi}_1^0}})/{m_{\tilde{\chi}_1^0}}$ and $\mathcal{R}_{\tilde{e}_R}\equiv({m_{\tilde{e}_R}-m_{\tilde{\chi}_1^0}})/{m_{\tilde{\chi}_1^0}}$.
		\label{table1}}
\end{table}

 %%%%%%%%%%%%%%%%%%%%%%%%%%%%%%%%%%%%%%%%%%%%%%%%%%%%%%%%%%%%
The most appealing feature of models with low-energy SUSY is the prediction of thermal relic DM. Indeed, a colorless, stable, neutral weekly interacting massive particle (WIMPs) leads to present DM density roughly agreeing with the observation \cite{Akrami:2018vks}. The SSMs, with the R parity conserved, provide a satisfactory theoretical framework for the existence of such particles.\cite{neutralinodarkmatter,darkmatterreviews}. In the MSSM, the LSP is a mixture of bino, neutral wino, and higgsinos mass eigenstates. Neutral wino and higgsino state couple directly to SM gauges bosons, and this is complicated by the fact that neutralinos may annihilate too many final states: $f\bar{f}$, $W^+W^-$, $ZZ$, $ZH$, $hh$, and including $H$, $A$, and $H^\pm$ heavy Higgs bosons states. Many processes contribute to each of these final states. It is useful to begin by considering the pure bino-like neutralino to investigate the bulk region to avoid resonance annihilation and coannihilation. Thus, we considered the $99.9\%$ bino-like neutralino to preclude large annihilation cross-section coming from the higgsino or wino components through the final gauge bosons state processes and three gauge bosons vertices that involve the hypercharge gauge bosons. The bino limit does not disappear for the process that has final $f \bar f$ states through a t-channel via sfermion exchange, i.e. bino DM always annihilates through the process $\tilde{\chi}_1^0 \tilde{\chi}_1^0\rightarrow f \bar{f}$ via t and u-channel sfermions exchange where $f, \bar{f}$ are standard model fermions. This process becomes efficient if the mass of the intermediate sfermions light. Therefore, the electroweak scale bino-like DM is a viable option in our study through the light right-handed sleptons mediating the annihilation, which has long been called the bulk region, the most natural version of neutralino DM, wherein no coannihilation or resonance annihilation mechanism is necessary to suppress the relic abundance to the cosmological-viable range. We also considered the 
$2m_{\tilde{\chi}_1^0}<<m_{H^0}, m_{A^0}$ and $2m_{\tilde{\chi}_1^0}>>m_{h}$ to avoid the A-funnel/Higgs resonance in our study, while considering the numerical calculation and analytical results, we employed $\mathcal{R}_{\tilde{\phi}}\gtrsim10\%$, where $\mathcal{R}_{\tilde{\phi}}\equiv({m_{\tilde{\phi}}-m_{\tilde{\chi}_1^0}})/{m_{\tilde{\chi}_1^0}}$ to ignore the portion proportional to coannihilation, because the ratio of a mass difference $\mathcal{R}_{\tilde{\phi}}$ is important instead of the absolute mass difference. Thus, to negligible the coannihilation processes we employ $\mathcal{R}_{\tilde{e}_R}\equiv({m_{\tilde{e}_R}-m_{\tilde{\chi}_1^0}})/{m_{\tilde{\chi}_1^0}}\gtrsim10\%$, and $\mathcal{R}_{\tilde{\tau}_1}\equiv({m_{\tilde{\tau}_1}-m_{\tilde{\chi}_1^0}})/{m_{\tilde{\chi}_1^0}}\gtrsim10\%$ as the first and second right-handed sleptons, $\tilde{\tau}_1$ and $\tilde{e}_R$ are naturally light. Therefore, our search imposes $\mathcal{R}_{\tilde{\tau}_1}>\mathcal{R}_{\tilde{e}_R}$ while varying the ratio $\mathcal{R}_{\tilde{e}_R}$, and vice versa. Numerical findings reveal that the ratio $\mathcal{R}_{\tilde{\tau}_1}\gtrsim10\%$ implies that $m_{\tilde{\chi}_1^0}\leq120.4$ GeV. In the latter case where we employ $\mathcal{R}_{\tilde{\tau}_1}>\mathcal{R}_{\tilde{e}_R}$ and varied the ratio $\mathcal{R}_{\tilde{e}_R}$, all points with an $\tilde{e}_R$ NLSP are excluded by the ATLAS soft lepton SUSY search \cite{ATLAS:2019lng}. Then, the only viable MSSM region from the GmSUGRA in the bulk is for the case $m_{\tilde{\chi}_1^0}<m_{\tilde{\tau}_1}<m_{\tilde{e}_R}=m_{\tilde{\mu}_R}$.  All points in Fig. (\ref{F1}) show $\mathcal{R}_{\tilde{\tau}_1}$ vs. neutralino graph, which satisfies the experimental constraints mentioned in the previous section. 

The $\tilde{\tau}_1$-$\tilde{e}_R$ plane represented in Fig.(\ref{F11}). Within the bulk region from the GmSUGRA, the upper limit of $\tilde{\tau}_1$ and $\tilde{e}_R$ around $138$ GeV and $270$ GeV respectively recognize that these right-handed sleptons and bino LSP are naturally light. Note that we want to comment here that Fermi-Lab Collaboration has recently announced the measurement of muon anomalous magnetic moment results from Run 2 and 3 \cite{Muong-2:2023cdq} which strongly suggests new physics around 1 TeV \cite{Ahmed:2021htr}. Fig (\ref{Fg-2}) shows the contribution of our bulk scenario to $\Delta a_{\mu}$ up to $1\sigma$ deviation from the central value. Though our red solutions contribute up to $1\sigma$ and $2\sigma$ to $\Delta a_{\mu}$ have been excluded by the LHC but the brown (under-saturated) and orange (over-saturated) solutions are safe. 

The SUSY has been searched extensively at the LHC. The $\pm1\sigma$ observed and expected exclusion limits from the recent ATLAS SUSY updated summary plots \cite{ATLAS:2023xco} of the \cite{ATLAS:2019lff,ATLAS:2019lng,ATLAS:2022hbt} search for direct pair production of sleptons and charginos decaying into final states with two leptons are shown in Fig. (\ref{F4}) with superimposed several benchmark points from the bulk region in the GmSUGRA. All points in Fig.(\ref{F4}) belong to traditional annihilation only and stick to our requirement $\mathcal{R}_{\tilde{\tau}_1}\gtrsim10\%$. The green points are $\tilde{\tau}_1$ and blue points are $\tilde{e}_R=\tilde{\mu}_R$ from the GmSUGRA. It is important to note that the thin orange-shaded region in the ATLAS graph in Fig. (\ref{F4}) corresponds to $\tilde{e}_R=\tilde{\mu}_R$ and not $\tilde{\tau}_1$ constraints. The green-shaded region depicts the $\tilde{\tau}_1$ constraints. Though this paper has not shown, that the situation is similar with respect to CMS SUSY searches for the EW production of sleptons\cite{CMS:2020bfa,CMS:2023qhl,CMS:2022syk}. Our analytical results in Fig (\ref{F4}) revealed that our bulk region is still beyond the LHC approach and because of the compressed nature of these spectra, this bulk may not be probed at the LHC. It is anticipated that these light sleptons could be probed when the forthcoming circular colliders, namely the Future Circular Collider (FCC-ee) \cite{FCC:2018byv,FCC:2018evy} at CERN and the Circular Electron-Positron Collider (CEPC) \cite{CEPCStudyGroup:2018ghi} power up their beams. Table-\eqref{table1} provides two benchmark points, highlighting our findings consistent with the unexplored LHC region under the constraints discussed in the previous section. 

	\begin{figure}[h!]
	\centering \includegraphics[width=7.90cm]{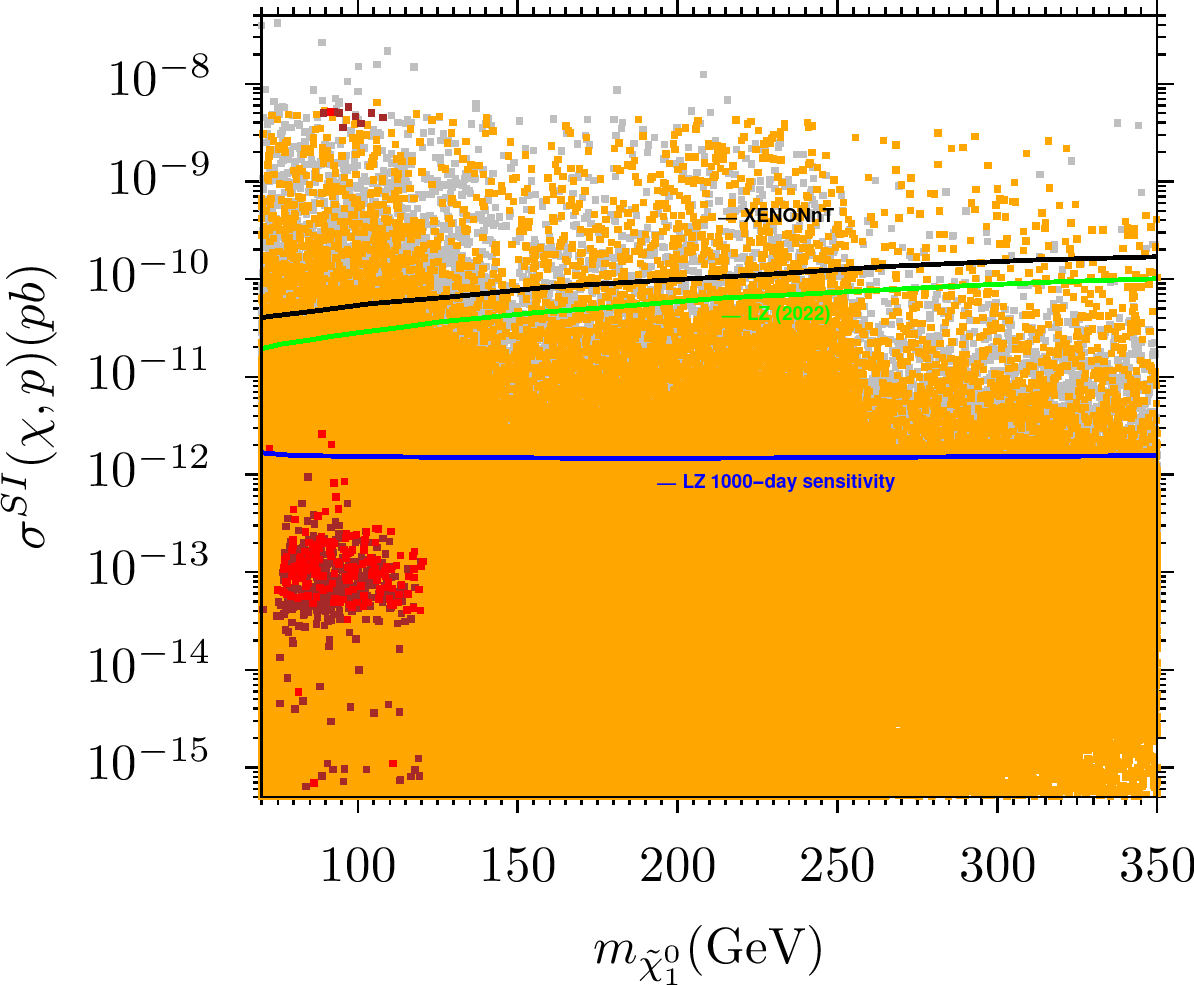}
	\centering \includegraphics[width=7.90cm]{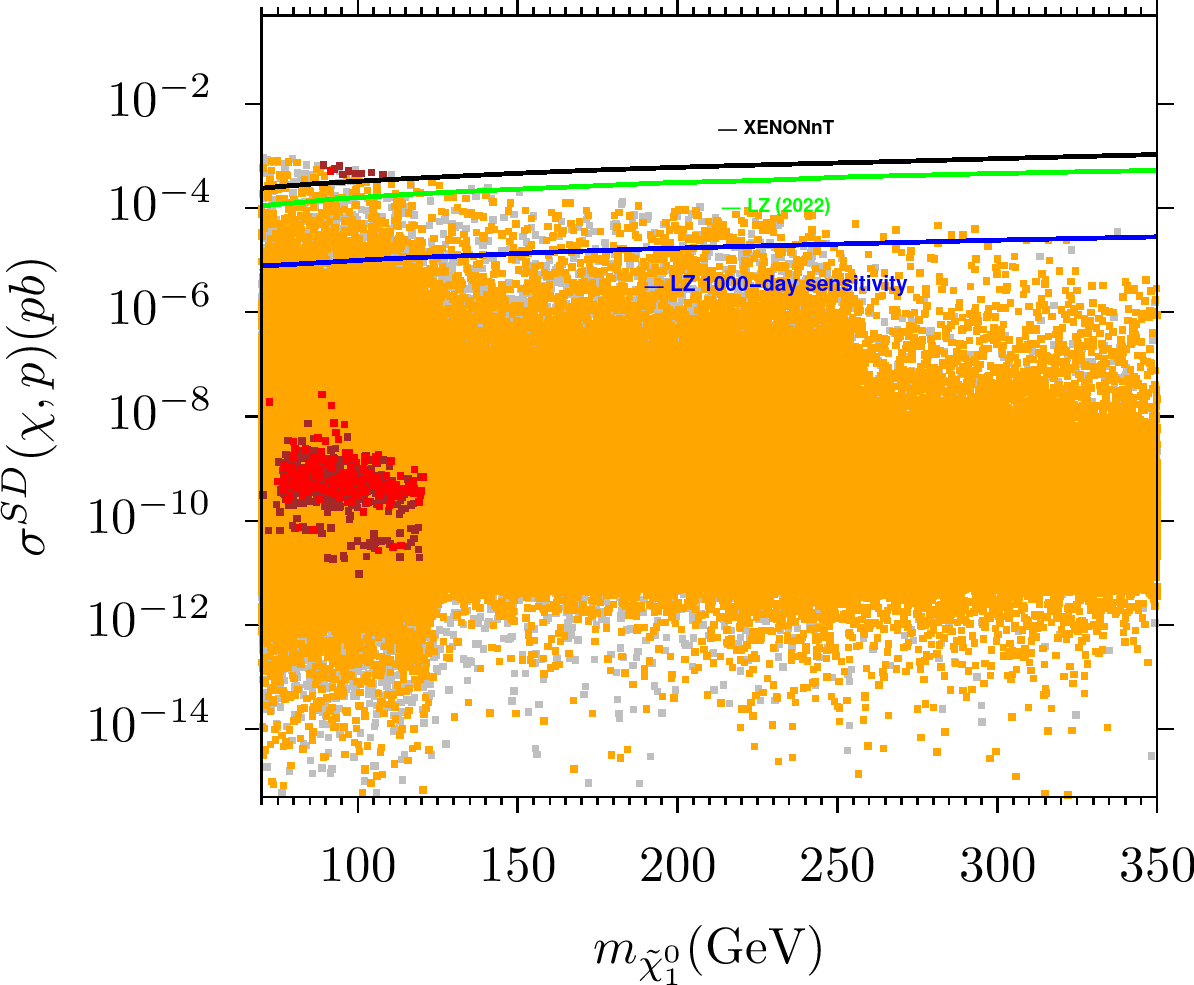}
	\caption{\small The color coding and the description are the same as in Fig. (\ref{F1}) with $\mathcal{R}_{\tilde{\tau}_1}\gtrsim10\%$. The spin-independent (top) and spin-dependent (bottom) neutralino-proton scattering cross-section vs. the neutralino mass in reference from XENONnT (solid blue line) \cite{XENON:2023cxc} and LUX-ZEPLIN (Solid green line LZ(2022) and black line LZ-1000 day sensitivity) \cite{LZ:2018qzl, LZ:2022lsv}}. 
	\label{F3}
\end{figure}

In Fig.\ref{F3}, we display the spin-independent (top) and spin-dependent (bottom) neutralino-proton scattering cross-section vs. the neutralino mass, adhering to our requirement $\mathcal{R}_{\tilde{\tau}_1}\gtrsim10\%$. In both panels, the solid blue line depicts the XENONnT \cite{XENON:2023cxc}, the green line shows the LUX-ZEPLIN(2022) \cite{LZ:2022lsv}, and the black line represents the 1000-day LUX-ZEPLIN experiment \cite{LZ:2018qzl}. Plots in the $m_{\tilde{\chi}_1^0}- \sigma_{SI}$ plane show that almost all of Planck2018 bound satisfy solutions are below the blue and green lines, except a handful of points. In the figure, we also notice that the 1000-day LUX-ZEPLIN experiment (black line) \cite{LZ:2018qzl} is anticipated to probe more of our GmSUGRA bulk. The plots in the $m_{\tilde{\chi}_1^0}- \sigma_{SD}$ plane depict that our solutions are consistent with the current research of the direct-detection experiments.

%%%%%%%%%%%%%%%%%%%%%%%%%%%%%%%%%%%%%%%%%%%%%%%%%%%%%%%%%%%%%%%%Table1%%%%%%%%%%%%%%%%%%%%%%%%%%%%%%%%%%%%%%%%%%%%%%%%%%%%%%%%%%%%%%%%%%%%%%%%%%%%%%%%%%%%%%%%%%%%%%%%%%%%%%%%%%%%%%%%%%%%

%%%%%%%%%%%%%%%%%%%%%%%%%%%%%%%%%%%%%%%%%%%%%%%%%%%%%%%

\section{conclusion}
\label{conclusion}

We have discussed the naturally generated light sfermions and thus a light bino LSP, a scenario we regard as the most natural DM or bulk region via the EWSUSY breaking from the GmSUGRA in the MSSM. We derive a region of parameter space that supports light right-handed sleptons and a light LSP, known as the bulk region, where $m_{\tilde{\chi}_1^0}\leq120.4$ GeV with negligible coannihilation correct with Plank2018 relic bound with a light stau NLSP and upper limits on $m_{\tilde{\tau}_1}$ and $m_{\tilde{e}_R}$ about $138$ GeV and $270$ GeV respectively. In particular, we open up the bulk region in MSSM via EWSUSY from the GmSUGRA that allows bino annihilation via t-channel slepton exchange, leading to "supersymmetric DM" at all with respect to DM. Our analytical results uncovered that the light right-handed sleptons in the bulk region could be beyond the LHC reach and may be probed in the forthcoming advanced era of circular colliders, the Future Circular Collider (FCC-ee) at CERN and the Circular Electron Positron Collider (CEPC). Furthermore, our DM implication is correct with XENONnT, LUX-ZEPLIN experiments, Planck2018 DM relic density bounds, and could be probed during the presently running 1000-day LUX-ZEPLIN experiment.

%%%%%%%%%%%%%%%%%%%%%%%%%%%%%%%%%%%%%%%%%%%%%%%%%%%%%%%%%%%%%%%%%%%%%%%
\textbf{Acknowledgments.--} TL is supported in part by the National Key Research and Development Program of China Grant No. 2020YFC2201504, by the Projects No. 11875062, No. 11947302, No. 12047503, and No. 12275333 supported by the National Natural Science Foundation of China, by the Key Research Program of the Chinese Academy of Sciences, Grant NO. XDPB15, by the Scientific Instrument Developing Project of the Chinese Academy of Sciences, Grant No. YJKYYQ20190049, and by the International Partnership Program of Chinese Academy of Sciences for Grand Challenges, Grant No. 112311KYSB20210012.

%%%%%%%%%%%%%%%%%%%%%%%%%%%%%%%%%%%%%%%%%%%%%%%%%%%%%%%%%%%%%%%%%%%%%%%

.

%%%%%%%%%%%%%%%%%%%%%%%%%%%%%%%%%%%%%%%%%%%%%%%%%%%%%%%%%%%%%%%%%%%%%%%

\end{document}